\documentclass[aps,prb,twocolumn,citeautoscript,superscriptaddress,showpacs]{revtex4}

\usepackage{graphicx}
\usepackage{dcolumn}
\usepackage{bm}
\usepackage{amssymb}
\usepackage[colorlinks=true,%
            citecolor=blue,%
            linkcolor=blue,%
            urlcolor=blue]{hyperref}

\begin{document}

\title{Electronic charge redistribution in LaAlO$_3$(001) thin films deposited at SrTiO$_3$(001) substrate: First principles analysis and the role of stoichiometry}

\author{Alexandre Sorokine}
\email{as08504@lu.lv}
\affiliation{Faculty of Physics and Mathematics, University of Latvia, 8 Zellu Str., Riga LV-1002, Latvia}
\affiliation{Institute for Solid State Physics, University of Latvia, 8 Kengaraga Str., Riga LV-1063, Latvia}
\author{Dmitry Bocharov}
\affiliation{Faculty of Physics and Mathematics, University of Latvia, 8 Zellu Str., Riga LV-1002, Latvia}
\affiliation{Faculty of Computing, University of Latvia, 19 Raina blvd., Riga LV-1586, Latvia}
\author{Sergei Piskunov}
\affiliation{Institute for Solid State Physics, University of Latvia, 8 Kengaraga Str., Riga LV-1063, Latvia}
\affiliation{Faculty of Computing, University of Latvia, 19 Raina blvd., Riga LV-1586, Latvia}
\author{Vyacheslavs Kashcheyevs}
\affiliation{Faculty of Physics and Mathematics, University of Latvia, 8 Zellu Str., Riga LV-1002, Latvia}
\affiliation{Faculty of Computing, University of Latvia, 19 Raina blvd., Riga LV-1586, Latvia}


\begin{abstract}
We present a comprehensive first-principles study of the electronic charge redistribution in atomically sharp LaAlO$_3$/SrTiO$_3$(001) heterointerfaces of both n- and p-types allowing for non-stoichiometric composition. Using two different computational methods within the framework of the density functional theory (linear combination of atomic orbitals and plane waves) we demonstrate that conducting properties of LaAlO$_3$/SrTiO$_3$(001) heterointerfaces strongly depend on termination of LaAlO$_3$(001) surface.  We argue that both the  ``polar catastrophe'' and the polar distortion scenarios may be realized depending on the interface stoichiometry. Our calculations predict that heterointerfaces  with a non-stoichiometric film---either  LaO-terminated n-type or AlO$_2$-terminated p-type---may exhibit the conductivity of n- or p-type, respectively, independently of LaAlO$_3$(001) film thickness.
\end{abstract}

\pacs{68.35.Ct, 68.35.Md, 73.20.At}

\maketitle

\section{Introduction}\label{sec:intro}

The discovery of conducting interfaces between two initially insulating materials---TiO$_2$-terminated (001) surface of SrTiO$_3$ (STO) substrate and LaAlO$_3$ (LAO) thin film deposited on top of it\cite{Ohtomo-Nature427p423}---has attracted strong scientific interest during the last few years.\cite{Hwang-NatMater2012-review,ADMA:ADMA200903800,0953-8984-22-4-043001,Chambers2010317,Bristowe-JPhys-2011} The high application potential of LAO/STO heterointerfaces has been demonstrated, e.g., by fabrication of highly voltage-tunable oxide diodes\cite{jany:183504} that utilize the advantage of the electric-field controlled interfacial metal--insulator transition of LAO/STO.\cite{Cen-natmat7p298,forg:053506}

Conductivity of an atomically flat interface in the limit of large film thickness can be understood \cite{Ohtomo-Nature427p423} from electrostatic considerations within the so-called ``polar catastrophe'' picture. From the perspective of formal charges, the atomic planes in [001] direction (which we refer to as monolayers) are neutral for STO (SrO$^0$ and TiO$_2^0$), but charged for LAO (LaO$^+$ and AlO$_2^-$).  Transition from STO to LAO can be p-type (from SrO$^{0}$ to AlO$_2^-$) or n-type (from TiO$_2^0$ to LaO$^+$). The corresponding jump of the surface charge at the interface would create an electric field  inside LAO increasing linearly with the distance from the interface---a ``polar catastrophe''. This bulk polarization of the LAO film can be compensated (thus averting the ``catastrophe'') if $0.5e$ per unit cell area is transferred from the LAO film surface onto the interface, resulting in a maximal sheet carrier density of $n=0.5/a^2=3.3\times 10^{14}$~cm$^{-2}$ (here $a=3.90\,$\AA{} is the lattice constant of STO assuming epitaxial matching of the LAO film).
This estimate is immune to dielectric relaxation and bond covalency/charge smearing effects~\cite{Bristowe-JPhys-2011} and thus provides a useful reference in the thick film limit.

For sufficiently thin films, however, the polar catastrophe may be tolerated \cite{PhysRevLett.102.107602} and a metal--insulator transition occurs \cite{PhysRevB.80.241107} as a function of the number of epitaxial monolayers of LAO deposited. From electrostatic perspective, in a sufficiently thin film the internal field does not develop a potential difference large enough to overcome the dielectric gap. The accumulation of oppositely charged monolayers leads to progressive band bending until the critical thickness (5~u.c. or 10 monolayers for n-type structures, according to experimental \cite{PhysRevB.80.241107} and theoretical \cite{PhysRevB.79.245411,PhysRevB.83.085408,li:013701} evidence) is reached beyond which the chemical forces are overcome, and the charge redistribution occurs. This mechanism is known as ``polar distortion'' \cite{PhysRevLett.102.107602,0953-8984-22-4-043001}. As far as it is known, p-type interfaces do not exhibit this mechanism, as covalent forces overwhelm electrostatic ones.

Similar electostatic arguments may be applied to non-stoichiometric LAO/STO structures, i.e. the ones with an odd number of LAO monolayers.  The LAO films in these structures possess one extra electron (or a hole for p-type interfaces) per unit cell area compared to the parent bulk material, thus they should be conducting (with  $n=1/a^2=6.6\times 10^{14}$~cm$^{-2}$) irrespective of the thickness or the presence of STO substrate.  However, as seen from the stoichiometric example, for thin films the competition between the semi-covalent bonds and long-range electrostatics is very sensitive to the the number of monolayers deposited.  The nature of the conducting layer in non-stoichiometric LAO/STO interfaces is the main subject of our \emph{ab initio} investigation.

 Recent experimental reports indicate that La/Al ratio in non-stoichiometric LAO films may be controlled during epitaxial growth \cite{PhysRevB.83.085408,droubay:124105}. Atomically sharp interfaces are produced by molecular beam epitaxy (MBE) in which thermal energies of evaporated incident ions are low, about 0.1~eV,
thus MBE avoids intermixing of cations at the interface\cite{PhysRevB.80.241107,Chambers2010317}.  However,
the vaporization process used to facilitate transfer through the vapor LAO phase does not guarantee preservation of the target stoichiometry\cite{Chambers2010317}, which makes  room for a possibility to control the film growth monolayer-by-monolayer.  We note that for different preparation methods of LAO/STO interfaces, e.g. pulsed laser deposition (PLD), other mechanisms may give rise to conductivity. One of proposed mechanisms is formation of the high density of oxygen vacancies, which are generated in the STO substrate while depositing LAO thin film and can be responsible for increase of sheet carriers density up to 5 $\times$ 10$^{17}$ cm$^{-2}$ for PLD-grown n-LAO/STO interfaces if the sample is not annealed.\cite{Ohtomo-Nature427p423,Basletic-natmat7p621} The insulating behavior of p-LAO/STO has been also ascribed to that the holes can be trapped by two electrons located at the oxygen vacancies created in the STO substrate.\cite{PhysRevB.79.161402}

Yet another scenario for LAO/STO interface conductivity that may take place in PLD-prepared structures is based on the suggestion that the La/Sr cation intermixing due to ion bombardment effect (inherent in PLD and post-growth treatment) may lead to the formation of one or two layers of metallic La$_{1-x}$Sr$_x$TiO$_3$.\cite{Chambers2010317,PhysRevLett.103.146101,Kalabukhov-EPL93p37001} The thermodynamical stability for intermixed configurations has been recently reported.\cite{0953-8984-22-31-312201,Chambers2010317}

In this paper, we aim to construct a clear picture of charge density redistribution both in stoichiometric and non-stoichiometric interfaces of either type and LAO film thickness from 1 to 11 monolayers (0.5--5.5~u.c.).  The \emph{ab initio} calculation methods employed are based on the density functional theory (DFT) using a hybrid exchange--correlation functional.  We contrast stoichiometric/non-stoichiometric and p-type/n-type structures utilizing identical methods and computation parameters.  The B3PW functional~\cite{becke-hybr} used in the \textsc{crystal} code~\cite{CRman2009} with atomic basis set (BS) contains a ``hybrid'' of the DFT exchange and correlation functionals with exact non-local Hartree--Fock (HF) exchange.  For comparison, the selected set of interface configurations has been also modeled using the Perdew--Wang generalized gradient approximation (PW91-GGA) density functional\cite{PW1,PW3} as implemented in the periodic plane-wave (PW) code \textsc{vasp}.\cite{VASP2011}

We find that covalent effects in non-stoichiometric films are less pronounced than in stoichiometric ones and the structures are metallic in accordance with formal charges considerations.  As Ti--O bond strength exceeds Sr--O bond strength by ca.~$120\,$kJ$\,$mol$^{-1}$ (obtained considering formation enthalpies for respective oxides), in p-type IFs (where SrO monolayer is at the IF) we can expect covalent forces to be stronger than in n-type IFs.  This, in its turn, leads to an approximately uniform free charge distribution through the film, whereas when covalent forces are weaker---as in n-type IFs---the free charge is forced to the edges of the LAO film (the surface and the IF) resulting in a bi-layered electron gas structure.

Experimental works show that stoichiometric p-type interfaces exhibit no measurable conductivity\cite{PhysRevB.80.241107}, but annealed stoichiometric n-type interfaces with LAO film thickness $\geqslant 5$~u.c. have free electron density in range 1--$3\times 10^{13}$~cm$^{-2}$\cite{PhysRevB.80.241107}.  Similar densities (2--$7\times 10^{13}$~cm$^{-2}$) are obtained from first principle calculations\cite{PhysRevB.79.245411,PhysRevB.83.085408,li:013701} (cf. $3.3\times 10^{14}$~cm$^{-2}$ predicted from electrostatic considerations).

The paper is structured as follows.  Section~\ref{sec:compdet} describes the computational details of our calculations. The main part of the paper is formed by Sec.~\ref{sec:Results}.  In Sec.~\ref{sec:lao} we give an estimate of the thermodynamic stability and discuss the electronic structure of ideal LaO- and AlO$_2$-terminated LAO(001) surfaces. Section~\ref{sec:heterostructures} presents electronic charge distribution for n-LAO/STO and p-LAO/STO heterointerfaces and discusses their relation to the experimental and computational data available in the literature.  Our conclusions are summarized in Section~\ref{sec:conc}.

\section{Computational details}\label{sec:compdet}

In this study LAO/STO heterointerfaces are modeled by means of two different methods: (i) linear combination of atomic orbitals (LCAO) within the framework of hybrid density functional approach, and (ii) PW calculations using the GGA density functional.

To perform hybrid LCAO calculations, we used the periodic \textsc{crystal} code\cite{CRman2009}, which employs Gaussian-type functions centered on atomic nuclei as the BSs for expansion of the crystalline orbitals. The BSs used in this study were taken from the following sources: For Sr, Ti and O in the form of 311d1G, 411d311dG, and 8-411d1G, respectively, from Ref.~\onlinecite{Piskunov-CMS29}; for Al in the form of 8-621d1G from Ref.~\onlinecite{0953-8984-19-39-395021}; for La in the form of 311-31d3f1 from \textsc{crystal}'s homepage~\cite{CRman2009} ($f$-type polarization Gaussian function with the exponent $\alpha=0.475$ has been added according to prescription given in Ref.~\onlinecite{piskunov:012410}). For Al and O all electrons are explicitly included. The inner core electrons of Sr and Ti are described by small-core Hay--Wadt effective pseudopotentials \cite{hw3}, while the non-relativistic pseudopotential of Dolg et al.\cite{Dolg-TCA75p173} was adopted for La.

We employ the hybrid B3PW exchange--correlation functional~\cite{becke-hybr} which accurately reproduces the basic bulk and surface properties of a number of ABO$_3$ perovskite materials\cite{Piskunov-CMS29,Piskunov-CMS41p195,Piskunov-SS575p75,Kotomin-PCCP10p4258}. The cutoff threshold parameters of \textsc{crystal} for Coulomb and exchange integrals evaluation (ITOL1--ITOL5) have been set to 7, 8, 7, 7, and 14, respectively. Calculations were considered as converged only when the total energy obtained in the self-consistency procedure differed by less than $10^{-7}$~a.u. in two successive cycles. Effective charges on atoms as well as net bond populations have been calculated according to the Mulliken population analysis.\cite{Mulliken-JCP23p1833,Mulliken-JCP23p1841,Mulliken-JCP23p2338,Mulliken-JCP23p2343}

As the second method the periodic total-energy code \textsc{vasp}\cite{VASP2011} based on the use of a PW BS was applied. The cut-off energy has been chosen to be 520 eV. The non-local GGA exchange--correlation functional Perdew--Wang-91 (PW91) was employed.\cite{PW1,PW3} Scalar relativistic projector augmented wave (PAW) pseudopotentials in our calculations contain 11 valence electrons ($5s^25p^65d^16s^2$) for La, 3 electrons ($3s^23p^1$) for Al, 10 electrons ($4s^24p^65s^2$) for Sr, 12 electrons ($3s^23p^63d^24s^2$) for Ti, and 6 electrons ($2s^22p^4$) for O, respectively.
Bader topological analysis\cite{bader} has been adopted to obtain net charges on atoms in \textsc{vasp} calculations.

In both \textsc{vasp} and \textsc{crystal} calculations the reciprocal space integration was performed by sampling the Brillouin zone with the $8\times8\times1$ Pack--Monkhorst mesh\cite{monkhorst} for all surface structures under consideration. For bulk computations we applied sampling with the $8\times8\times8$ Pack--Monkhorst mesh. Such samplings provide balanced summation in direct and reciprocal lattices.

\begin{table}
\caption{\label{tab:LAObulk} Calculated equilibrium lattice constants ($a_0$ in \AA), atomic net charges ($Q_{\rm atom}$ in e), cation--O bond populations ($P_{\rm A/B-O}$ in milli $e$), and band gaps ($\delta$ in eV) of bulk LAO and STO in their high-symmetry $Pm\bar{3}m$ cubic phase. Shown are data obtained by means of both hybrid B3PW and standard GGA PW91 functionals. Negative bond population means atomic repulsion. Last two columns contain available experimental results for comparison.}
\begin{ruledtabular}
\begin{tabular}{lcccccc}
                               & LAO    & LAO    & STO    & STO    & LAO    & STO    \\
                               & {\scriptsize (B3PW)} & {\scriptsize (PW91)} & {\scriptsize (B3PW)} & {\scriptsize (PW91)} & {\scriptsize (Exp.)} & {\scriptsize (Exp.)} \\
             \hline
             $a_0$             & 3.802  &  3.808     & 3.910      & 3.918      & 3.811\cite{PhysRevB.72.054110} & 3.905\cite{abramov} \\
             $Q_{\rm La/Sr}$   & 2.43   &  2.14      & 1.87       & 1.60       & --         & --         \\
             $Q_{\rm Al/Ti}$   & 2.07   &  3.00      & 2.35       & 2.10       & --         & --         \\
             $Q_{\rm O}$       & $-1.50$  &  $-1.78$     & $-1.41$      & $-1.23$      & --         & --         \\
             $P_{\rm La/Sr-O}$ & 4      &  --        & $-10$        & --         & --         & --         \\
             $P_{\rm Al/Ti-O}$ & 152    &  --        & 88         & --         & --         & --         \\
             $\delta$          & 5.51   &  3.18      & 3.64       & 1.77       & 5.6\cite{lim:4500} & 3.25\cite{benthem} \\
\end{tabular}
\end{ruledtabular}
\end{table}

Taking into account that STO substrate at room temperature possesses perfect cubic structure, in our study we treat both LAO and STO in their high symmetry $Pm\bar{3}m$ cubic phase. In fact, the bulk crystal structure of LAO, having space group $R\bar{3}c$ (rhombohedral) with $a_0 = 5.364$~\AA{} and $c_0 = 13.108$~\AA{} at room temperature,\cite{PhysRevB.72.054110} can be represented by a pseudocubic unit cell with $a_0 = 3.790$~\AA. At 821~K the structure of LAO transforms to become cubic with $a_0 = 3.811$~\AA.\cite{PhysRevB.72.054110} Though the heterointerface assumes the transition between two intrinsically different crystal symmetries: $Pm\bar{3}m$ the substrate and $R\bar{3}c$ in the film, whereby thin films are expected to adapt to the substrate.\cite{PhysRevB.85.045401}

Table~\ref{tab:LAObulk} lists main bulk properties for both crystals. We note that the band gaps obtained by means of hybrid B3PW computation scheme are in better agreement with experimentally observed results. Therefore in this paper we mainly discuss the results obtained by means of B3PW while results obtained using PW91 functional are published for comparative purposes in order to make our study consistent with earlier \emph{ab initio} calculations performed basically on LDA- or GGA-DFT ground.

Surface structures were modeled using a single slab model for LCAO calculations and a multi-slab model with vacuum gap of 20~\AA{} for PW calculation. To compensate the dipole moment arises at charged surfaces, our slabs are symmetrically terminated. STO substrate contains 11 alternating (SrO)$^0$ and (TiO$_2$)$^0$ atomic monolayers, while from 1 to 11 alternating (LaO)$^+$ and (AlO$_2$)$^-$ atomic monolayers were used for LAO film of the LAO/STO interface. Coordinates of all atoms in the LAO/STO heterointerfaces were allowed to relax. Due to symmetry constrains atomic displacements were allowed only along $z$-axis. Taking into account that the mismatch of $\sim$2.5\% between LAO and STO lattice constants arises during LAO epitaxial growth, in our modeling we have allowed relaxation of their joint lattice constant to minimize the strain effect. 

\section{Results and discussion}\label{sec:Results}

\subsection{LAO(001) surfaces}\label{sec:lao}

Before general discussion of LAO/STO interfaces studied here, in this subsection we provide a comprehensive description of electronic and thermodynamic properties of both LaO- and AlO$_2$-terminated pristine LAO(001) thin films.

\subsubsection{
Electronic properties}\label{sec:atelprop}

Pristine LAO(001) thin films were modeled using symmetrical 9-monolayer slab model. Considering formal ionic charges, LAO(001) has alternating (LaO)$^+$ and (AlO$_2$)$^-$ surface monolayers and can be either LaO- or AlO$_2$-terminated surface. Both LaO- and AlO$_2$-terminations are studied. La/Al excess ratio is 1.25 and 0.8 for LaO- and AlO$_2$-terminated LAO(001) films, respectively. Monolayers in LAO(001) possess a net charge, the repeat slab unit cell has a non-zero dipole moment and therefore LAO(001) is type III polar surface according to Tasker's classification.\cite{Tasker-JPCSSP12p4977} This means, that perfect and unreconstructed ($1\times 1$) LAO(001) surfaces considered here can be stabilized by transferring of a half an electron (or hole) from the surface to the slab body that normally results in atomic and electronic reconfiguration at the surface.

\begin{table}
\caption{\label{tab:LAOsurfDQ} Calculated deviations in surface monolayer net charge ($\Delta Q$ in $e$), and deviations of cation--O bond populations ($\Delta P_{\rm A/B-O}$ in milli $e$) in corresponding atomic monolayer relative to the bulk values (see Table~\ref{tab:LAObulk}). Shown are data obtained by means of hybrid B3PW exchange--correlation functional. Surface monolayers are numbered beginning from the center of the slab (0 means the central monolayer of the symmetrical slab unit cell).}
\begin{ruledtabular}
\begin{tabular}{ccccccc}
                               & \multicolumn{3}{c}{LaO-term.}    & \multicolumn{3}{c}{AlO$_2$-term.}   \\
             No.       & M-layer & $\Delta Q$ & $\Delta P_{\rm A/B-O}$ & M-layer & $\Delta Q$ & $\Delta P_{\rm A/B-O}$ \\
             \hline
             4                 & LaO    &  $-0.32$     & 10         & AlO$_2$    & 0.46       & 100        \\
             3                 & AlO$_2$&  $-0.02$     & $-16$        & LaO        & $-0.02$      & $-4$         \\
             2                 & LaO    &  $-0.09$     & 0          & AlO$_2$    & 0.02       & $-10$        \\
             1                 & AlO$_2$&   0.00     & $-2$         & LaO        & $-0.02$      & $-4$         \\
             0                 & LaO    &  $-0.05$     & $-2$         & AlO$_2$    & 0.02       & $-10$        \\
\end{tabular}
\end{ruledtabular}
\end{table}

In Table~\ref{tab:LAOsurfDQ} we list the changes in surface (LaO)$^+$ and (AlO$_2$)$^-$ monolayer net charges with respect to their bulk values (see Table~\ref{tab:LAObulk}). Due to partly covalent nature of La--O and Al--O bonds (positive $P_{\rm A/B-O}$ in Table~\ref{tab:LAObulk}) net charges of La, Al, and O deviate from their formal ionic values of $+3$, $+3$, and $-2$, respectively. The La--O hybridization between La $5d$ and O $2p$ states lead to atomic charges of $2.43e$, $2.07e$, and $-1.50e$ for La, Al, and O, respectively. As a result, LaO and AlO$_2$ monolayers possess a bulk monolayer charge of $\pm 0.93e$ instead of formal ionic $\pm 1e$ charge. According to the Table~\ref{tab:LAOsurfDQ} surface monolayer of LaO-terminated LAO(001) attracts 0.32 electrons, while other monolayers of the slab get the rest of 0.14 electrons to compensate the surface polarity. On the contrary, surface monolayer of AlO$_2$ terminated LAO(001) solely receives 0.46 holes. Covalency of surface La--O bond is only slightly increased (bond population increased only by 10 milli $e$), while calculated covalency of surface Al--O bond is practically two times larger than in the bulk, that, to some extent, may compensate relatively modest surface relaxation of AlO$_2$-terminated LAO(001) with respect to LaO-terminated one.

\begin{figure}
\includegraphics[width=8.5cm]{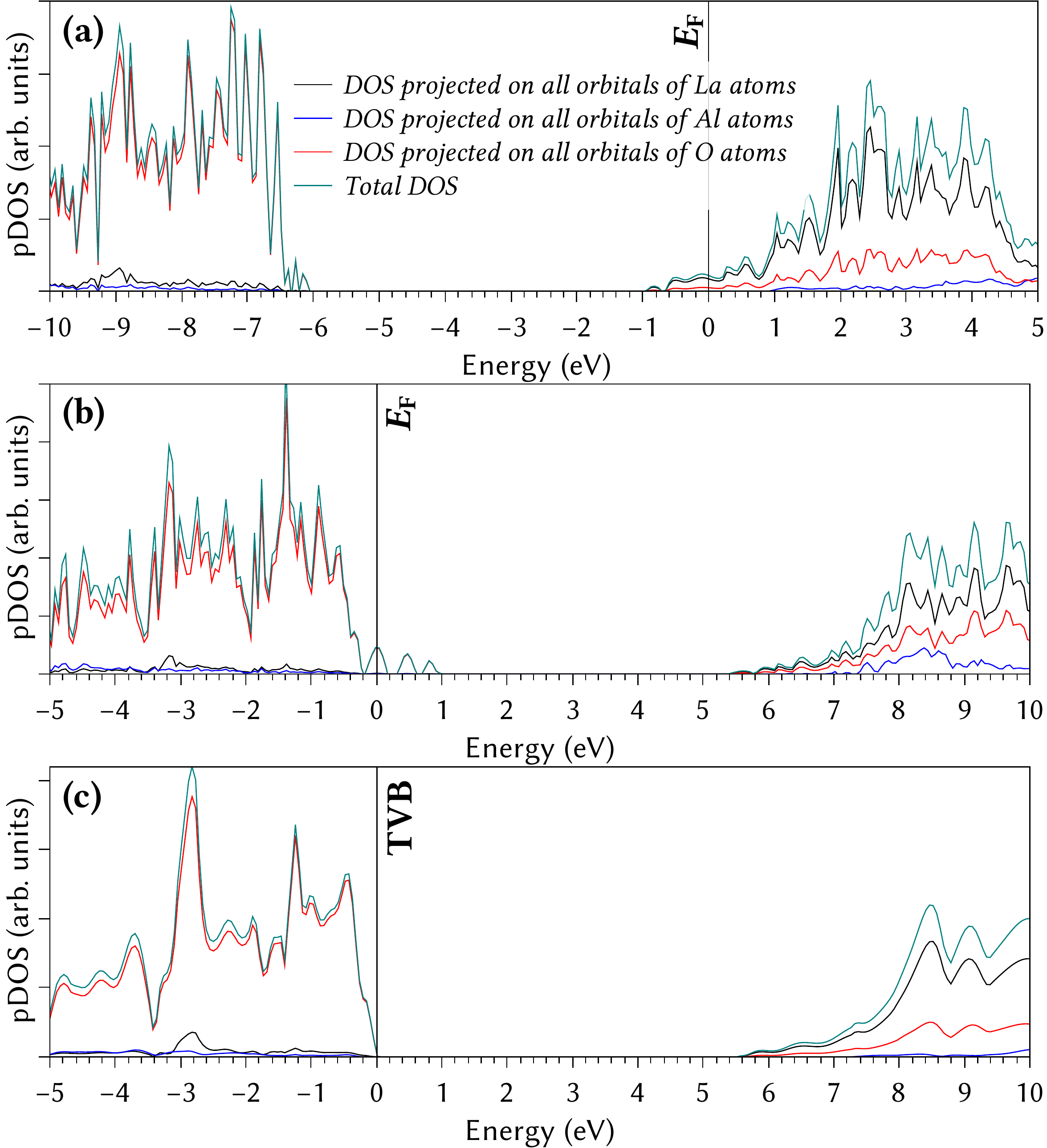}
\caption{\label{fig:DOSLAO} (Color online) Projected density of states  as calculated by means of B3PW hybrid exchange--correlation functional: (a) LaO-terminated LAO(001), (b) AlO$_2$-terminated LAO(001), (c) LAO bulk.  TVB stands for the top of valence band.}
\end{figure}

Fig.~\ref{fig:DOSLAO} shows the density of states (DOS) projected onto all orbitals of La, Al, and O atoms of LAO bulk and both LaO- and AlO$_2$-terminated LAO(001) surfaces as well. In case of LAO bulk (Fig.~\ref{fig:DOSLAO}c) the top of valence band is formed by O $2p$ orbitals, while the bottom of conduction band is formed mainly by La $5d$ states. La--O hybridization is well pronounced. Calculated band gap of 5.51 eV is in excellent agreement with its experimental value of 5.6 eV.\cite{lim:4500} In case of LaO-terminated surface (Fig.~\ref{fig:DOSLAO}a) gained excess of electrons shifts the Fermi level up to unoccupied level that gives raise to electron conductivity. In its turn the AlO$_2$-terminated surface (Fig.~\ref{fig:DOSLAO}b) experiences the lack of electrons that shifts Fermi level down to valence band and thus reveals the existence of hole conductivity.

\subsubsection{Thermodynamic stability}\label{sec:TD}

The thermodynamic formalism adopted in the current study to estimate the stability of both LaO- and AlO$_2$-terminated LAO(001) surfaces has been thoroughly described in Refs.~\onlinecite{piskunov:121406} and~\onlinecite{heifets:115417} (see also references therein). The stable crystalline surface has to be in equilibrium with both LAO bulk and surrounding oxygen atmosphere assuming that an exchange of atoms between surface and environment is allowed. Therefore, the most stable surface has the lowest Gibbs free surface energy defined as
\begin{eqnarray}
\Omega_{\text{t}}(T,p)=&&\frac{1}{2A}[E^{slab}_t-N_{\text{Al}}E^{\rm LAO}_{\rm bulk}-
(N_{\text{La}}-N_{\text{Al}})\Delta\mu_{\text{La}} \nonumber \\
&&-(N_{\text{O}}-3N_{\text{Al}})\Delta\mu_{\text{O}}(T,p)], \label{eq:Omega}
\end{eqnarray}
where $t$ indicates the surface terminations, $A$ the unit cell surface area, $N_i$ the number of atoms of type $i$ in the slab unit cell, ${E^{slab}_t}$ is the total energy of a slab with $t$ surface terminations and $E^{\text{LAO}}_{\rm bulk}$ is the LAO total energy averaged per five-atom perovskite unit cell. $\Delta\mu_i=\mu_i-E^i_{\rm bulk}$, (${i}={\text{La,Al}}$) are deviations of chemical potentials for metal atoms from their energy in the bulk metals. For the oxygen atom such a deviation is considered with respect to the energy of an oxygen atom in the ground triplet state of an O$_2$ molecule $\Delta\mu_{\text{O}}=\mu_{\text{O}}-\frac{1}{2}E^{\text{O}_2}$. Because $pV$ term ($V$ is unit
cell volume) and the differences in vibrational Gibbs free energy between the bulk solid and a corresponding slab is negligibly small,\cite{Reuter-PRB65} we omit these two contributions. This permits replacing the Gibbs free energies in Eq. ({\ref{eq:Omega}) and in the following formul\ae{} with the total energies obtained from \emph{ab initio} calculations.

In order to avoid the precipitation of relevant metals and oxides at LAO surface, as well as to prevent metal atoms to leave the sample the following conditions must be satisfied:
\begin{eqnarray}
&{0>\Delta\mu_{\text{La}}}, \ \ \ {0>\Delta\mu_{\text{Al}}},& \label{eq:bound-La-Al}    \\
&{E^f_{\text{LaAlO}_3}}-{E^f_{\text{Al}_2\text{O}_3}}<{2\Delta\mu_{\text{La}}+3\Delta\mu_{\text{O}}}<
{E^f_{\text{La}_2\text{O}_3}},& \label{eq:bound-La2O3-Al2O3}
\end{eqnarray}
where $E^f_n$ is the formation energies of material $n$ listed in Table~\ref{tab:Form-Energies}.

\begin{table}
\caption{\label{tab:Form-Energies} Formation energies per formula
unit used in analysis of surface stability. Experimental values are
taken from Ref.~\onlinecite{NIST-JANAF}.}
\begin{ruledtabular}
\begin{tabular}{ldd}
             Material     & \multicolumn{1}{c}{$E^{f}$, eV}  &
\multicolumn{1}{c}{Exp. $E^{f}$, eV}  \\
             \hline
             La$_2$O$_3$  &    -17.52    &      -18.64       \\
             Al$_2$O$_3$  &    -16.68    &      -17.37       \\
             LaAlO$_3$    &    -17.68    &                   \\
\end{tabular}
\end{ruledtabular}
\end{table}

We evaluate the oxygen chemical potential $\Delta\mu_{\text{O}}(p,T)$ as a function of partial gas pressure and temperature using the standard experimental thermodynamical tables\cite{NIST-JANAF} as it was done in Refs.~\onlinecite{Reuter-PRB65,heifets:115417}. $\Delta \mu_{\text{O}}(T,p_{\text{O}_2})$ is the variation of oxygen chemical potential due to temperature and pressure of the surrounding oxygen atmosphere. In addition to the experimental variation it contains a correction term $\delta\mu^0_{\text{O}}=0.03$ eV, which compensates the difference between the experimentally determined variation of the oxygen chemical potential and the reference state in current theoretical calculations (see Refs.~\onlinecite{doi:10.1021/nn800210v} and~\onlinecite{doi:10.1021/jp909401g} for a thorough discussion).

\begin{figure*}
\includegraphics[width=16.0cm]{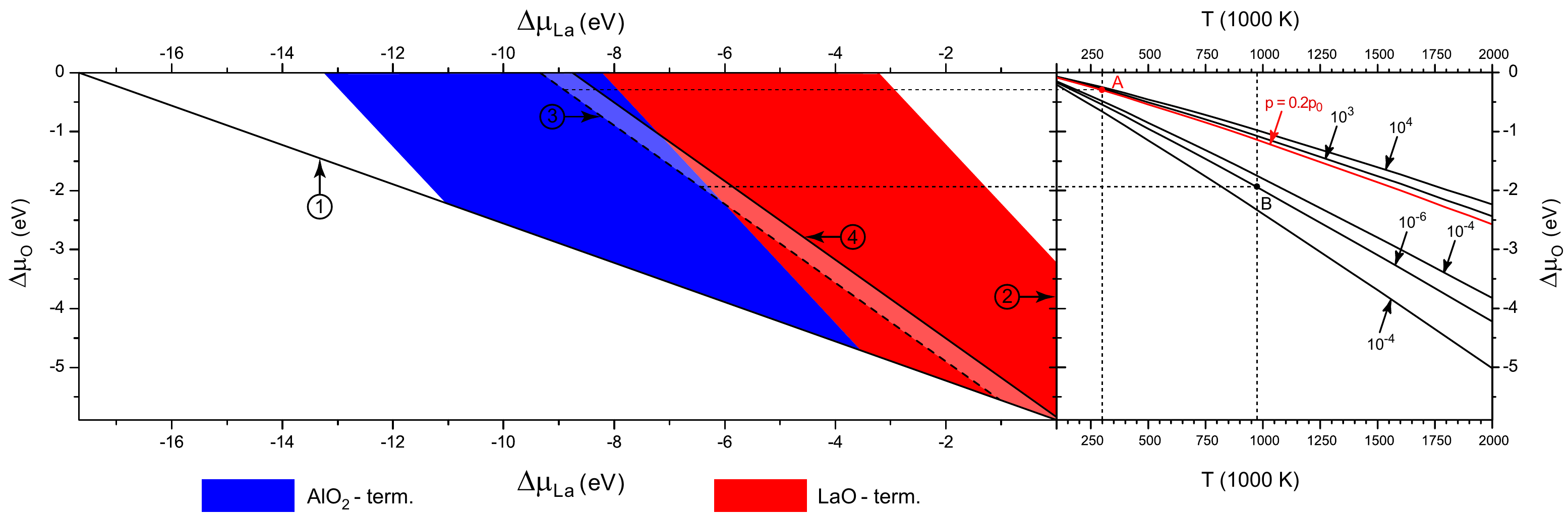}
\caption{\label{fig:TD-triang} (Color online) Thermodynamic stability
diagram as a function of O and La chemical potentials built for both LaO- and AlO$_2$-terminated
LAO(001) surfaces. Diagram contains precipitation conditions for both La and Mn metals, as well as for their trivalent oxides (La$_2$O$_3$ and Al$_2$O$_3$). Stable region is shown as
shaded area between La$_2$O$_3$ and Al$_2$O$_3$ precipitation
lines. The numbers from 1 to 4 in the circles indicate segregation lines for precipitation of:
1. Al, 2. La, 3. La$_2$O$_3$, 4. Al$_2$O$_3$. The right side shows a family of oxygen chemical potentials
under different conditions. The label $m$ indicates the O$_2$ gas
partial pressure: $10^m$ mbar. Red (gray) line corresponds to oxygen
partial pressure $p=0.2 p_0$ as in the ambient atmosphere. Point A stands for room temperature and ambient oxygen pressure, point B stands for typical temperature and pressure during LAO/STO(001) synthesis.}
\end{figure*}

Based on Eqs.~\ref{eq:Omega}, \ref{eq:bound-La-Al}, and \ref{eq:bound-La2O3-Al2O3}, the thermodynamic stability diagram is plotted in Figs.~\ref{fig:TD-triang}, showing the regions of stability of pristine LAO(001) surfaces with respect to precipitation of La$_2$O$_3$ and Al$_2$O$_3$ oxides. Fig.~\ref{fig:TD-vs-T} shows the thermodynamic stability diagram along the lines corresponding to precipitation of La$_2$O$_3$ and Al$_2$O$_3$ oxides as a function of $\Delta\mu_{\text{O}}$ related to the temperature scale at an oxygen pressure typical during LAO/STO synthesis ($P$ = 10$^{-6}$ mbar). To make such a diagram possible, according to prescription given in Ref.~\onlinecite{PhysRevB.83.073402} we replaced $\Delta\mu_{\text{La}}$ by
\begin{equation}\label{Eq:uLa1}
 \Delta\mu_{\text{La}}=\frac{1}{2}(E^{\rm f}_{\rm La_2O_3}-3\Delta\mu_{\rm O}),
\end{equation}
that corresponds to precipitation of La$_2$O$_3$ (lines 3 in Fig.~\ref{fig:TD-vs-T}) and by
\begin{equation}\label{Eq:uLa2}
 \Delta\mu_{\text{La}}=E^{\rm f}_{\rm LaAlO_3}-\frac{1}{2}(E^{\rm f}_{\rm Al_2O_3}-\frac{3}{2}\Delta\mu_{\rm O}),
\end{equation}
that corresponds to precipitation of Al$_2$O$_3$ (lines 4 in Fig.~\ref{fig:TD-vs-T}). Formation energies for oxides are taken from the Table~\ref{tab:Form-Energies}.

\begin{figure}
\includegraphics[width=8.0cm]{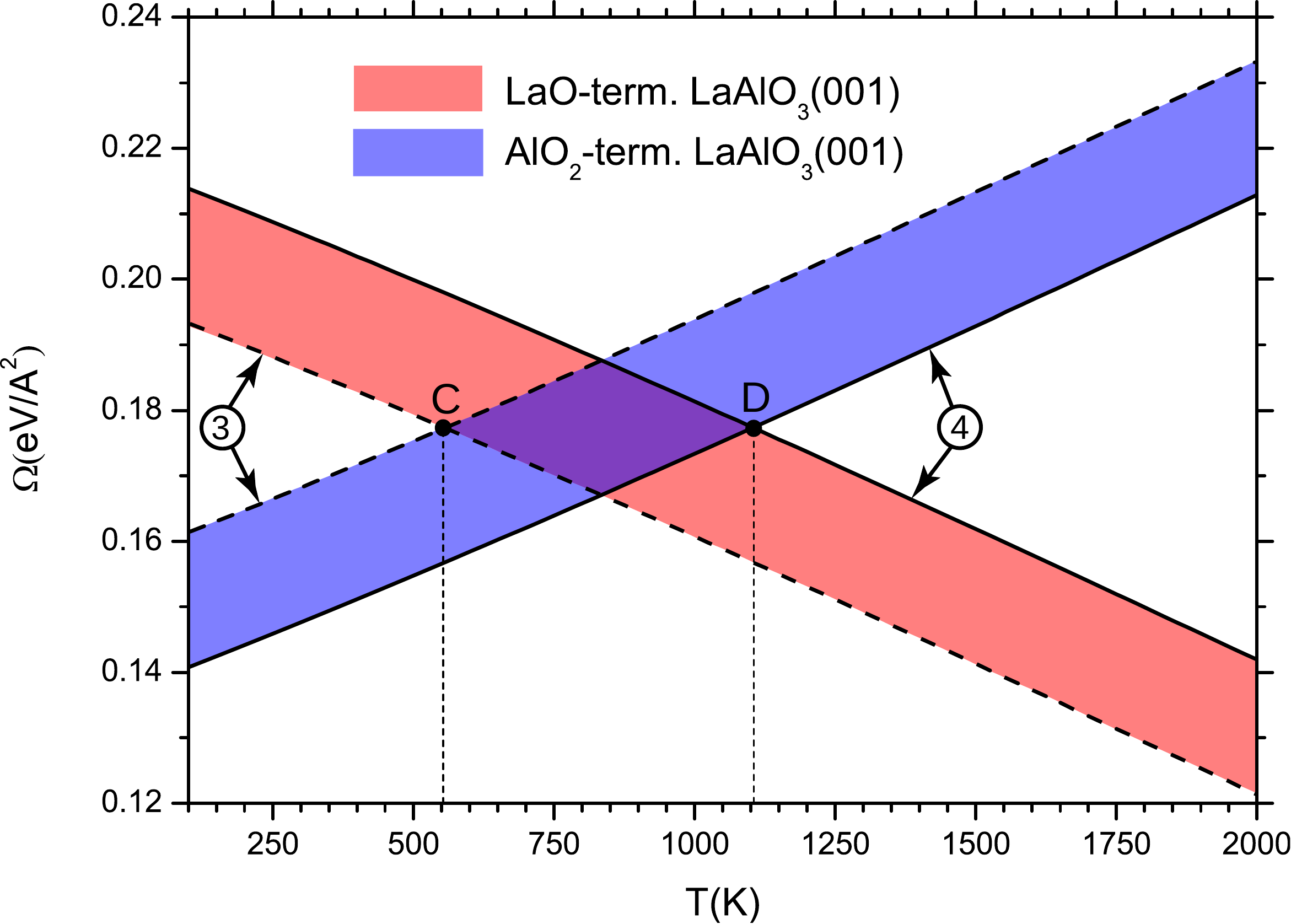}
\caption{\label{fig:TD-vs-T} (Color online) The thermodynamic stability diagram calculated along the La$_2$O$_3$ and Al$_2$O$_3$ precipitation lines (number 3 and 4 in the circles, respectively) with  $\Delta\mu_{\text{La}}$ defined according to Eqs.~\ref{Eq:uLa1} and \ref{Eq:uLa2}. The dependence on the oxygen chemical potential is converted to the appropriate temperature scale at an oxygen pressure typical during LAO/STO(001) synthesis ($P$ = 10$^{-6}$ mbar). The interval between points C and D correspond to temperature range where both LaO- and AlO$_2$-terminated LAO(001) surfaces are stable and may coexist.}
\end{figure}

From the calculated thermodynamic stability diagrams we can predict that at ultra-high vacuum (UHV) conditions typical during PLD synthesis of LAO/STO interfaces and low temperatures ($T<550$ K) the most stable is AlO$_2$-terminated surface, while at elevated temperatures ($T>1100$ K) stabilizes LaO-terminated surface. Between these temperatures both surface terminations may coexist. Further lowering of oxygen pressure shifts down these demarcated temperatures. This our prediction is in good qualitative agreement with time-of-flight scattering and recoiling spectrometry (TOF-SARS), atomic force microscopy (AFM), and photoelectron spectroscopy (PES) study performed by Rabalais and co-workers.\cite{10.1063.1.475535,vanderHeide1998350} They found that at temperatures less than 423 K, the surface is exclusively terminated by an Al-O layer, while at temperatures above 523 K the surface is exclusively terminated by a La-O layer. Between 423 K and 523 K  surface stoichiometry changed from AlO$_x$ to LaO$_x$ and thus mixed terminations were proposed. Moreover this change was found to be fully reversible. Rabalais and co-workers suggested that the surface termination change was caused by the formation of surface oxygen vacancies at high temperature, which drives the migration of the La atom to the surface and the Al atom into the bulk. More recent experimental study based on X-ray crystal truncation rod (CRT) analysis\cite{PhysRevB.64.235425} demonstrates that LAO(001) possesses Al-terminated structure at both room and high (670 K) temperatures with no evidence for the reversal of surface termination or for the formation of surface oxygen vacancy. Authors of Ref.~\onlinecite{PhysRevB.64.235425} explain the observation of La-rich termination in ion-scattering experiments\cite{10.1063.1.475535,vanderHeide1998350} by the effect of the increasing access to the lanthanum atom because of considerable surface oxygen relaxation that leads to a significant enhancement of the lanthanum atom signature. On the other hand, Marx and co-workers have observed the La-terminated LAO(001) with stoichiometry of (VLa$_4$O$_5$)$^{-0.5}$, where V is the lanthanum cation vacancy, i.e., each surface La is coordinate to four surface oxygens and four oxygens in the subsurface layer.\cite{PhysRevLett.98.086102} Therefore one may conclude that the experimental analyses have been performed at various conditions and report either LaO- and AlO$_2$-terminated LAO(001) or mixture of them, so it is not clear if surfaces reached thermodynamic equilibrium or not.

\emph{Ab initio} thermodynamical stability diagrams previously calculated for LAO(001) shows that LaO-terminated surface is more stable with respect to AlO$_2$-terminated one\cite{Tang2007149} and LaO-terminated surface containing oxygen vacancy is more stable than oxygen deficient AlO$_2$-terminated LAO(001) as well.\cite{1674-1056-17-2-049} Mixed surfaces with LaO- and AlO$_2$-terminations were not predicted. In fact, our thermodynamic analysis does not support this prediction.  From our point of view the main reason for such a discrepancy may be the different computational approach, DFT within local density approximation, used by Authors of Refs.~\onlinecite{1674-1056-17-2-049}, \onlinecite{Tang2007149}.

\subsection{LAO/STO heterointerfaces}\label{sec:heterostructures}
\subsubsection{Charge redistribution and electronic properties}\label{sec:chargeel}
%
%
Calculations of 
electronic properties of the LAO/STO(001) heterointerfaces were carried out using the symmetrically terminated slab model. The STO(001) substrate consisted of 11 atomic monolayers and could be terminated with either (TiO$_2$) monolayer in n-type heterostructures or with (SrO) monolayer in p-type heterostructures. Then monolayer-by-monolayer epitaxial growth was modeled adding a pair of respective monolayers of LAO(001) symmetrically to both sides of a substrate slab until deposited LAO(001) thin film reach thickness of up to 11 monolayers. In such way we construct 22 heterostructures of both types and of different LAO film thickness to model.  Note that 11-monolayer thick substrate and 20~\AA{} thick vacuum gap used for \textsc{vasp} GGA calculations is enough to avoid undesirable interaction of neighboring surfaces/interfaces and allows us to reach the equilibrium charge density redistribution in heterointerfaces under study.  Due to the restrictions by imposed symmetry, in our calculations atomic positions of all the heterointerfaces under study were relaxed along the $z$ axis. 

If we consider atomic displacements, we can see that cations and anions in LAO monolayers have considerably different displacements, thus electric dipole moment appears and accumulates within the thin film. Stoichiometric heterointerfaces have greater displacement differences between anions and cations than non-stoichiometric ones in LAO monolayers, while the situation is diametrically opposite for the STO monolayers. As we shall see further, the dipole moment creates an electric field, and its potential strongly correlates with the distortion of the band edges (so-called polar distortion), which then gives rise to the conductivity in stoichiometric LAO/STO(001) heterointerfaces of n-type.
%


\begin{figure}
\includegraphics[width=8.0cm]{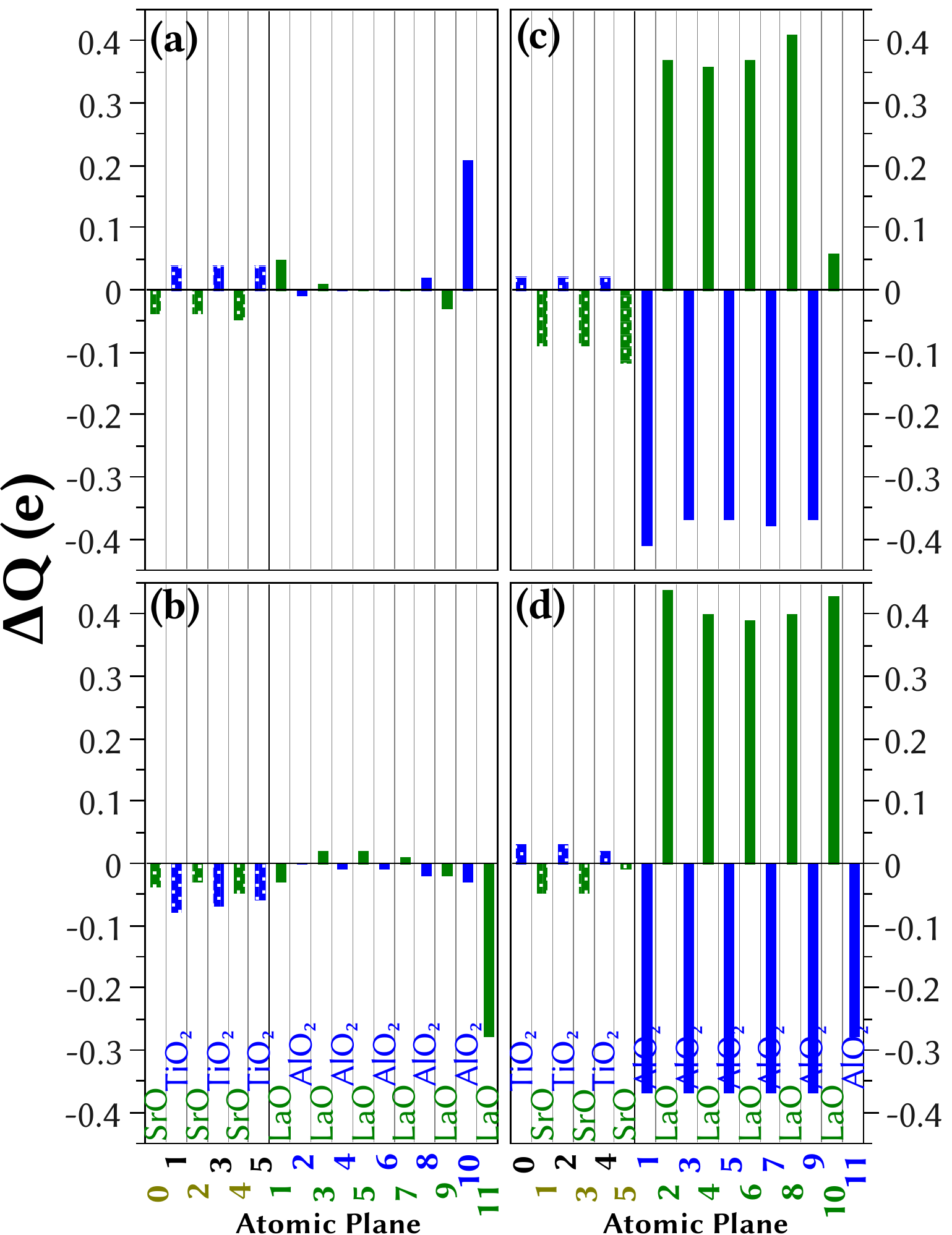}
\caption{\label{fig:Qpap} (Color online) Calculated deviations of Mulliken effective charge densities ($\Delta P_Q$) in AO- and BO$_2$-monolayers of (a, b) n-LAO/STO(001) and (c, d) p-LAO/STO(001) heterostructures with respect to charge densities in AO- and BO$_2$-monolayers of STO and LAO bulk, correspondingly. Calculations are performed using B3PW hybrid exchange-correlation functional. The $x$ axis shows the atomic monolayers from which atoms are originated. STO and LAO monolayers are numbered starting from the center of slab (0 means the central monolayer of the symmetrical slab unit cell). Monolayers (planes) are numbered separately for STO(001) substrate and for LAO(001) nanofilm. Panels (a, c) show charge density deviation for $N_{\rm LAO}=10$, while panels (b, d)---for $N_{\rm LAO}=11$.}
\end{figure}

To predict the charge redistribution in heterointerfaces we calculated the changes of net atomic Mulliken charges in comparison with the bulk phase of the LAO and STO parent materials. These charge deviations are shown in Fig.~\ref{fig:Qpap}a--d for LAO/STO(001) heterointerfaces of n- and p- type. From these one can clearly see, that deviation of charges are relatively small in the inner monolayers of the LAO film in n-type LAO/STO(001), not exceeding $0.03e$, whereas the same layers in the p-type LAO/STO(001) show quite large charge deviations $\pm (0.35$--$0.40)e$ from the parent bulk, and these are negative for AlO$_2$ monolayers and positive for LaO monolayers.

In both n- and p-type interfaces charges on the substrate monolayers did not vary substantially. For stoichiometric n-type and non-stoichiometric p-type interfaces these are about $\pm 0.04e$ for TiO$_2$ and SrO, respectively.  On the other hand, stoichiometric p-type interfaces show a small positive deviation of TiO$_2$ monolayer charges (ca.~$0.01e$) and about ten times bigger negative charge deviation for SrO monolayers.  Charge shifts in the substrates of stoichiometric n-type structures are all negative, and SrO shifts (ca.~$0.04e$) are smaller than TiO$_2$ shifts of ca.~$0.06e$.

Most significant deviations in atomic charges of n-type structures are located in the top-most monolayer---$+0.2e$ for stoichiometric structures and $-0.25e$ for non-stoichiometric ones---due to the surface effects and thus compensate the ``polar catastrophe'' as proposed from a pure ionic model\cite{Nakagawa-natmat5p204}. In p-type structures charge shifts in the surface layers are less pronounced than in the inner layers of the film and are $+0.05e$ and $-0.27e$ for LaO- and AlO$_2$-terminated structures, respectively.

Here charge redistribution only in the thickest structures investigated is shown. Respective graphs for thinner structures can be found in Refs. \onlinecite{SuppMater-n,SuppMater-p}.

\begin{figure*}
\includegraphics[width=14.0cm]{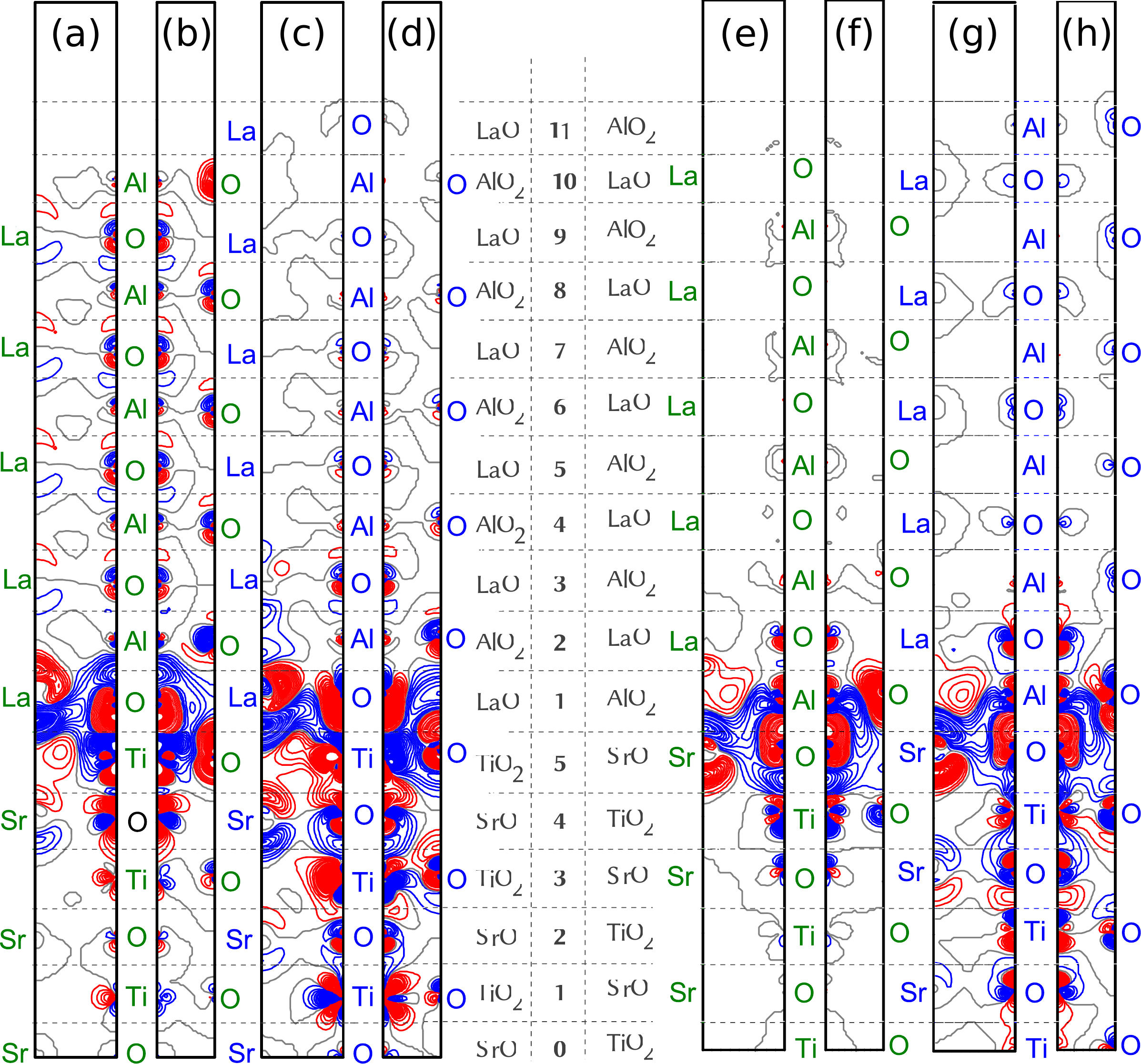}
\caption{\label{fig:ECHD} (Color online) Difference electron charge density maps calculated for (a--d) n-LAO/STO(001) and (e--h) p-LAO/STO(001) heterostructures: (a, e) (110) cross-section for $N_{\rm LAO}=10$, (b, f) (100) cross-section for $N_{\rm LAO}=10$, (c, g) (110) cross-section for $N_{\rm LAO}=11$, (d, h) (100) cross-section for $N_{\rm LAO}=11$. Red (dark gray), blue (light gray) and gray isolines describe positive, negative and zero values of the difference charge density, respectively. Isodensity curves are drawn from --0.025 to +0.025~$e$ \AA$^{-3}$ with an increment of 0.0005~$e$ \AA$^{-3}$. Right-side bar shows the atomic monolayers from which atoms are originated. Calculations are performed using B3PW hybrid exchange-correlation functional. STO and LAO monolayers are numbered beginning from the center of slab (0 means the central monolayer of the symmetrical slab unit cell). Monolayers (planes) are numbered separately for STO(001) substrate and for LAO(001) nanofilm.}
\end{figure*}

Another way to look at the problem of charge redistribution is to calculate, what happens with the electronic charge density in the heterostructures, compared to the isolated LAO and STO slab parts.  Charge density redistribution is defined as the electronic density in the heterointerface minus the sum of electron densities in separately isolated STO(001) substrate and LAO(001) thin film slabs and is depicted in Fig.~\ref{fig:ECHD} for both n- and p-type LAO/STO(001) interfaces.

These plots show us that the most significant distortions occur at the interface due to the compensation of the surface effects of the slabs. They also show that the electronic structure of the substrate of non-stoichiometric heterostructures is distorted stronger than that of stoichiometric ones. The situation in the thin films is opposite. This fact correlates with the argument in the section on atomic structure.

More illustrative property to consider is the polarization of all four of stoichiometric and non-stoichiometric n- and p-type heterointerfaces, which was already briefly introduced. It allows us to explain certain phenomena, such as the polar distortion, as well as to provide a mechanism for a partial compensation of the ``polar catastrophe''.

Let us assume that each one of considered interfaces possesses no net charge, thus it can be divided into multiple neutral slabs normal to $z$, in which net charge is also zero and average polarization of such slabs can be calculated.  Charge density function that should be used in the calculations, is estimated as if the charge of each atom $A$ is uniformly distributed over the plane $z=z_A$, reducing the task to one dimension.  Thus the projection of polarization vector on $z$-axis can be calculated as
\begin{equation}
\bar{P}_i = \frac{\sum_A z_A Q_A}{\Delta z},
\label{eqn:pols}
\end{equation}
$Q_A$ is the charge on atom $A$, $\Delta z$ is the thickness of the neutral allocated slab, to which the atom $A$ belongs and summation is performed over all the atoms in the $i$-th neutral slab.

In order to divide the interface in neutral slabs, it sometimes is necessary to split one monolayer's charge: One part of it compensates the remaining charge of the previous slab and the remainder goes to the next one.

\begin{figure}
\includegraphics[width=0.9\columnwidth]{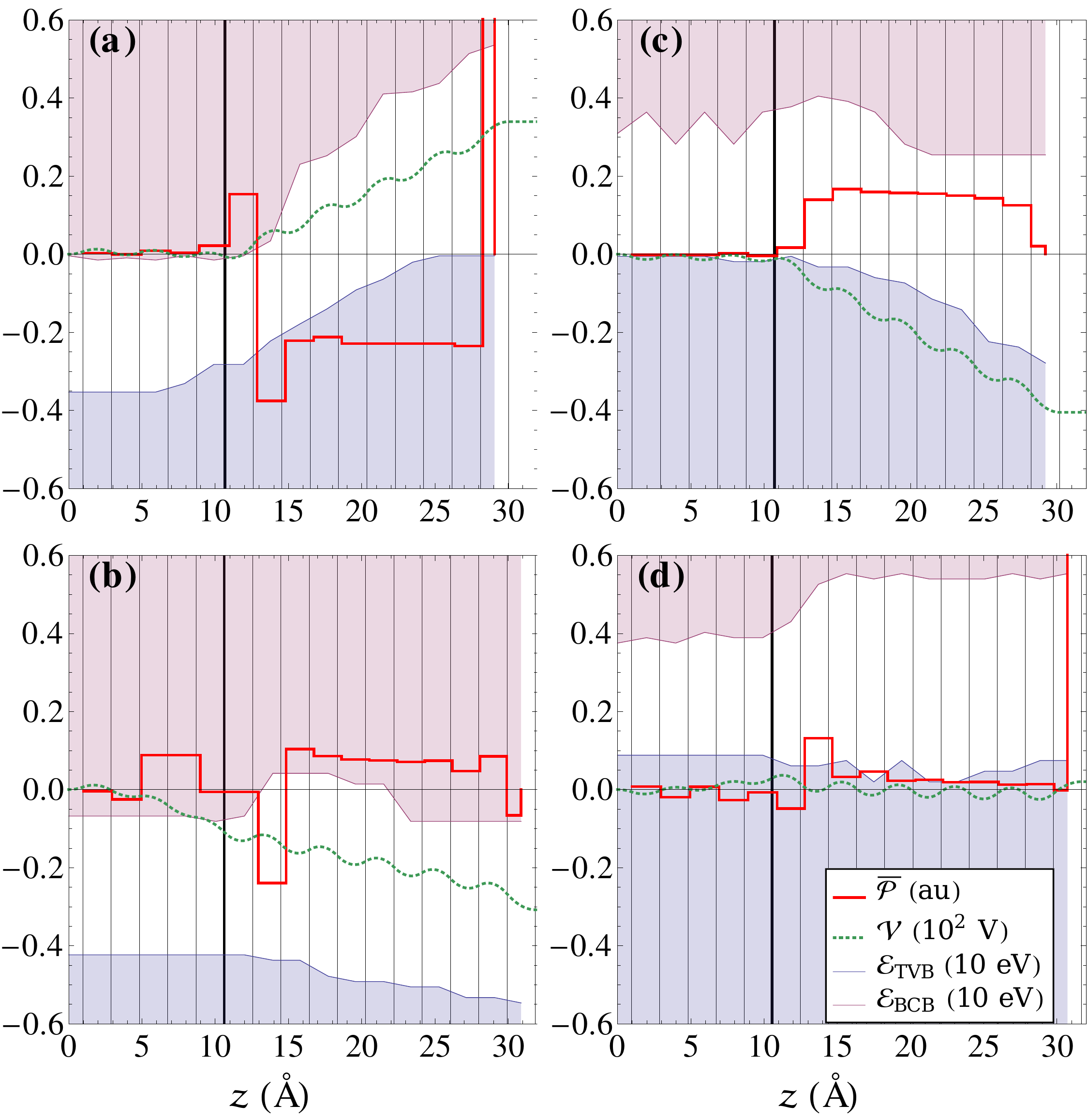}
\caption{\label{fig:Pavg} (Color online) Polarization (as calculated using Eq. (\ref{eqn:pols})), band edges and electrostatic potential of (a, b) n-LAO/STO(001) and (c, d) p-LAO/STO(001) heterostructures with (a, c) $N_{\rm LAO}=10$ and (b, d) $N_{\rm LAO}=11$ LAO monolayers. Zero at the energy scale corresponds to the Fermi level. Distances are measured from the central monolayer of the symmetrical slab unit cell.  TVB stands for the top of valence band, BCB stands for the bottom of conduction band.}
\end{figure}

The resulting polarization function $\bar P(z)$ is averaged using the moving average function, and the results for n- and p-type interfaces are shown together with the energies of band boundaries $E_{\rm TVB}$ and $E_{\rm BCB}$ and the potential due to intrinsic electrostatic field $V$ in Fig.~\ref{fig:Pavg}.  Here one can see, that LAO films of stoichiometric interfaces are strongly polarized, giving rise to the polar distortion of band edges.  On the other hand, there is rather weak LAO polarization in the non-stoichiometric interfaces meaning a weak polar distortion as is observed. The substrate is polarized more in non-stoichiometric case, which corresponds to Figs.~\ref{fig:ECHD}c,d and Figs.~\ref{fig:ECHD}g,h.  AlO$_2$-terminated structures possess substantial polarization in the top-most monolayer.  Top-most layer's polarization of LaO-terminated structures, on the other hand, is negligible. The interface monolayers of n-type structures are substantially polarized.

\begin{table}
\caption{\label{tab:n-gap-n} Band gaps ($\delta$ in eV) or sheet carrier density ($n_s$ in 10$^{14}$ cm$^{-2}$) of n-LAO/STO heterointerfaces as calculated by means of hybrid B3PW and PW91 exchange-correlation functionals. $N^{\rm LAO}_{\rm tot}$ stands for the total number of LAO(001) monolayers deposited atop STO(001) substrate.}
\begin{ruledtabular}
\begin{tabular}{cccccc}
                             &             & \multicolumn{2}{c}{B3PW (\textsc{crystal})} & \multicolumn{2}{c}{PW91 (\textsc{vasp})} \\
             $N^{\rm LAO}_{\rm tot}$ & Term. m-layer & $\delta$ & $n_s$  & $\delta$ & $n_s$ \\
             \hline
             1  & LaO     & --   & 6.04 & --   &      \\
             2  & AlO$_2$ & 3.65 & --   & 1.41 & --   \\
             3  & LaO     & --   & 6.07 & --   &      \\
             4  & AlO$_2$ & 2.91 & --   & 1.03 & --   \\
             5  & LaO     & --   & 5.91 & --   &      \\
             6  & AlO$_2$ & 1.96 & --   & 0.40 & --   \\
             7  & LaO     & --   & 6.20 & --   &      \\
             8  & AlO$_2$ & 1.07 & --   & 0.03 & --   \\
             9  & LaO     & --   & 6.27 & --   &      \\
             10 & AlO$_2$ & --   & 1.56 & --   & 0.16 \\
             11 & LaO     & --   & 6.13 & --   & 0.54 \\
\end{tabular}
\end{ruledtabular}
\end{table}

\begin{table}
\caption{\label{tab:p-gap-n} The same as Table~\ref{tab:n-gap-n}, but for p-LAO/STO(001) heterostructures.}
\begin{ruledtabular}
\begin{tabular}{cccccc}
                             &             & \multicolumn{2}{c}{B3PW (\textsc{crystal})} & \multicolumn{2}{c}{PW91 (\textsc{vasp})} \\
             $N^{\rm LAO}_{\rm tot}$ & Term. m-layer & $\delta$ & $n_s$ & $\delta$ & $n_s$ \\
             \hline
             1  & AlO$_2$ & --   & 6.65 & --   &      \\
             2  & LaO     & 4.00 & --   & 1.60 & --   \\
             3  & AlO$_2$ & --   & 7.27 & --   &      \\
             4  & LaO     & 4.05 & --   & 1.69 & --   \\
             5  & AlO$_2$ & --   & 9.08 & --   &      \\
             6  & LaO     & 4.05 & --   & 1.51 & --   \\
             7  & AlO$_2$ & --   & 7.90 & --   &      \\
             8  & LaO     & 3.80 & --   & 0.48 & --   \\
             9  & AlO$_2$ & --   & 6.97 & --   &      \\
             10 & LaO     & 2.92 & --   & 0.25 & --   \\
             11 & AlO$_2$ & --   & 10.2 & --   & 0.12 \\
\end{tabular}
\end{ruledtabular}
\end{table}

Electronic properties in a more experimentally measurable way can be represented as band gaps for insulating structures or as the concentration of charge carriers for conductors. These data obtained with \textsc{crystal} and \textsc{vasp} are represented in Tables~\ref{tab:n-gap-n} and \ref{tab:p-gap-n} for n- and p-type structures, respectively. Firstly, one can see that all the non-stoichiometric interfaces are conducting and free charge concentration is roughly equal within a type and does not depend on the LAO film thickness. p-Type structures possess greater carrier density than n-type structures, though experiments never showed conductive behavior in the former.

For stoichiometric structures insulating behavior is the default one. The thickness of the band gap decreases with the thickness of the LAO film both for n- and p-type structures. This eventually leads to the closing of the gap for the n-type interfaces with $N_{\rm LAO}\geqslant 10$ monolayers, which is in a good accordance with experimental works.\cite{PhysRevB.80.241107} The gap-diminishing tendency is less pronounced for the p-type structures and thus they are not found conducting at any thickness within this study.

The results obtained with \textsc{vasp} are given for qualitative comparison. They showed out to be in accordance with \textsc{crystal} results, but due to the specifics of the non-hybrid functional band gaps and free charge concentrations are far too small. Taking into account that the largest difference between calculated using \textsc{crystal} code and experimentally observed band gap of bulk materials is 0.39 eV (see Table \ref{tab:LAObulk}) we note that our \textsc{crystal} calculations give plausible results comparing to experimental data.

The total band gap described above gives us some valuable data on conducting--insulating behavior of the interfaces of different types. Nevertheless, it does not give us much information about the origin of conductivity. Thus it is more worthy to look at the positions of the band edges in energy scale separately for each monolayer. Such a decomposition is depicted in aforementioned Figures~\ref{fig:Pavg}a,b and \ref{fig:Pavg}c,d for n- and p-type structures, respectively. From these plots one can see, that band edges for stoichiometric interfaces are distorted, besides such a distortion leads to n-type conductivity in n-type structures thick enough and might hypothetically lead to the p-type conductivity in thicker p-type structures than investigated. Non-stoichiometric interfaces show little or no polar distortion, but it is not necessary for the appearance of the conductivity, because such structures contain non-stoichiometric LAO films, which are already conducting on their own. Our prediction on conductivity of non-stoichiometric LaO-terminated n-type LAO/STO(001) interface is in agreement with a recent  theoretical study performed by Pavlenko and Kopp (See Ref.~\onlinecite{Pavlenko20111114}) in which they show that LaO-terminated n-type LAO/STO(001) interface is metallic.

\section{Summary and concluding remarks}\label{sec:conc}

We have performed large-scale first-principles calculations on a number of both stoichiometric and non-stoichiometric LAO/STO(001) heterostructures. Two different \emph{ab initio} approaches have been applied: LCAO with hybrid B3PW and PW with PW91 exchange--correlation functionals within DFT.  Consistently within both approaches we predict that there exists a distortion in energies of band edges for stoichiometric structures which eventually leads to the appearance of the conductivity at a critical thickness in n-type interfaces or to the reduction of the band gap for p-type interfaces.  Non-stoichiometric interfaces were found to be conducting independently of the LAO film thickness and possessing little or no distortion of band edges. The conductivity appears due to the non-stoichiometry of the thin film which is a conductor on its own, as we demonstrate by a separate analysis of an isolated film.

The degree of distortion of the band edges agree well with the estimates of the internal electric field generated by changes in the atomic charges and the geometric relaxation of the atomic structure. We confirm these factors as the ones responsible for the rise of conductivity in stoichiometric n-type heterostructures. Calculated concentration of the free charge in the interfaces roughly agrees with the experimental data, being somewhat underestimated.

For non-stoichiometric n-type interfaces electron gas structure is monolayered with uniform distribution over both the film and the substrate, while for p-type interfaces it is bilayered with one part of free charge carriers located on $3d$ orbitals of Ti at the IF, while the other is located on La orbitals at the surface.  The total calculated $n\approx 6\times 10^{14}$~cm$^{-2}$ well accords with that, predicted from electrostatic assumptions $n_{\rm ES}=1/a^2=6.6\times 10^{14}$~cm$^{-2}$.  Of that the IF gas layer gets $n_{\rm IF}\approx 1.3\times 10^{14}$~cm$^{-2}$ and the surface gas layer gets $n_{\rm S}\approx 4.7\times 10^{14}$~cm$^{-2}$.

Thermodynamic analysis that we have performed for the pristine LAO(001) surface reveals that its both LaO- and AlO$_2$-terminations may co-exist at temperatures above 550 K. If LAO/STO(001) heterointerface is covered by LaO monolayer, charge compensation mechanism of deposited polar non-stoichiometric LAO film leads to the tendency of Ti$^{3+}$ formation at the interface (see Fig. \ref{fig:Qpap}). To some extend it may explain the unexpected observation of Ti$^{3+}$ photoemission spectroscopy peak from n-type LAO/STO interfaces grown at 873 K.\cite{PhysRevB.84.245124}

In general, we conclude that one should not disregard the stoichiometry aspect when considering ways to make the LAO/STO interfaces conducting as non-stoichiometric interfaces possess unique quasi-2D electron gas structure that gives an overall 2 times greater free charge carrier density in comparison with stoichiometric interfaces.  For stoichiometric n-type structures covalent and electrostatic forces' interplay leads to metal--insulator transition at critical film thickness, but for non-stoichiometric---to formation of bilayered (n-type IFs) or monolayered (p-type IFs) quasi-2D electron gas.

\begin{acknowledgments}
This work has been supported through the ESF project Nr.2009/0216/1DP/1.1.1.2.0/09/APIA/VIAA/044. The authors are thankful to R.~Evarestov, A.~Shluger, E.~Kotomin, Yu.~Purans, E.~Heifets, Yu.~Zhukovskii, J.~Timoshenko and P.~Nazarov for stimulating discussions.
\end{acknowledgments}

\bibliographystyle{apsrev4-1}
\bibliography{Xbib}

\appendix
\vspace{\stretch{1}}
\section{Supplementary Materials}
\begin{figure*}
\centering
\includegraphics[width=15cm]{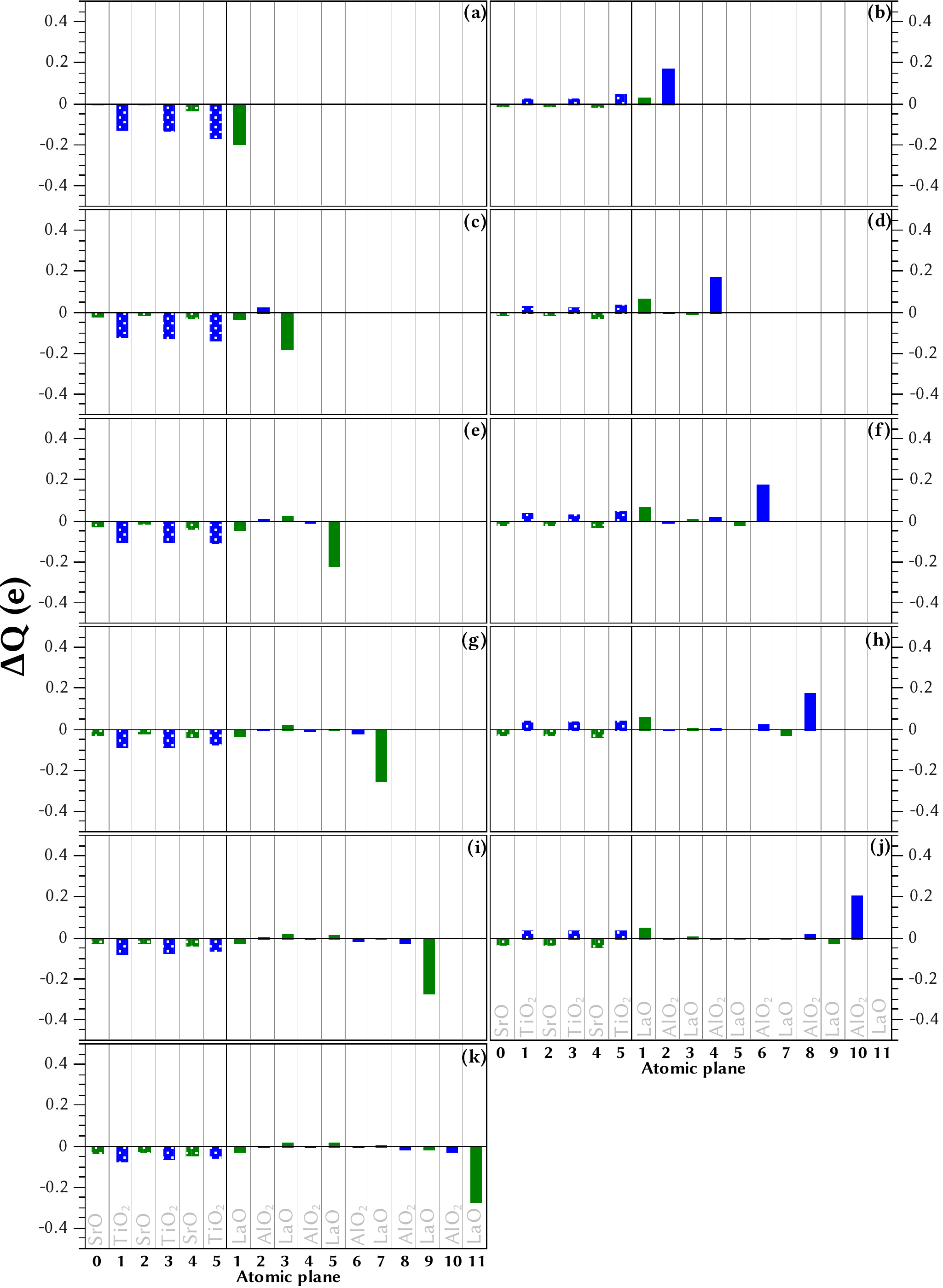}
\caption{\label{fig:Qnsupp} (Color online) Calculated deviations of Mulliken effective charge densities ($\Delta P_Q$) in AO- and BO$_2$-monolayers of n-LAO/STO(001) heterostructures with respect to charge densities in AO- and BO$_2$-monolayers of STO and LAO bulk, correspondingly. Calculations are performed using B3PW hybrid exchange-correlation functional. The $x$ axis shows the atomic monolayers from which atoms are originated. STO and LAO monolayers are numbered starting from the center of slab (0 means the central monolayer of the symmetrical slab unit cell). Monolayers (planes) are numbered separately for STO(001) substrate and for LAO(001) nanofilm. Number of monolayers increases from 1 to 11 for panels (a) to (k).}
\end{figure*}

\begin{figure*}
\centering
\includegraphics[width=15cm]{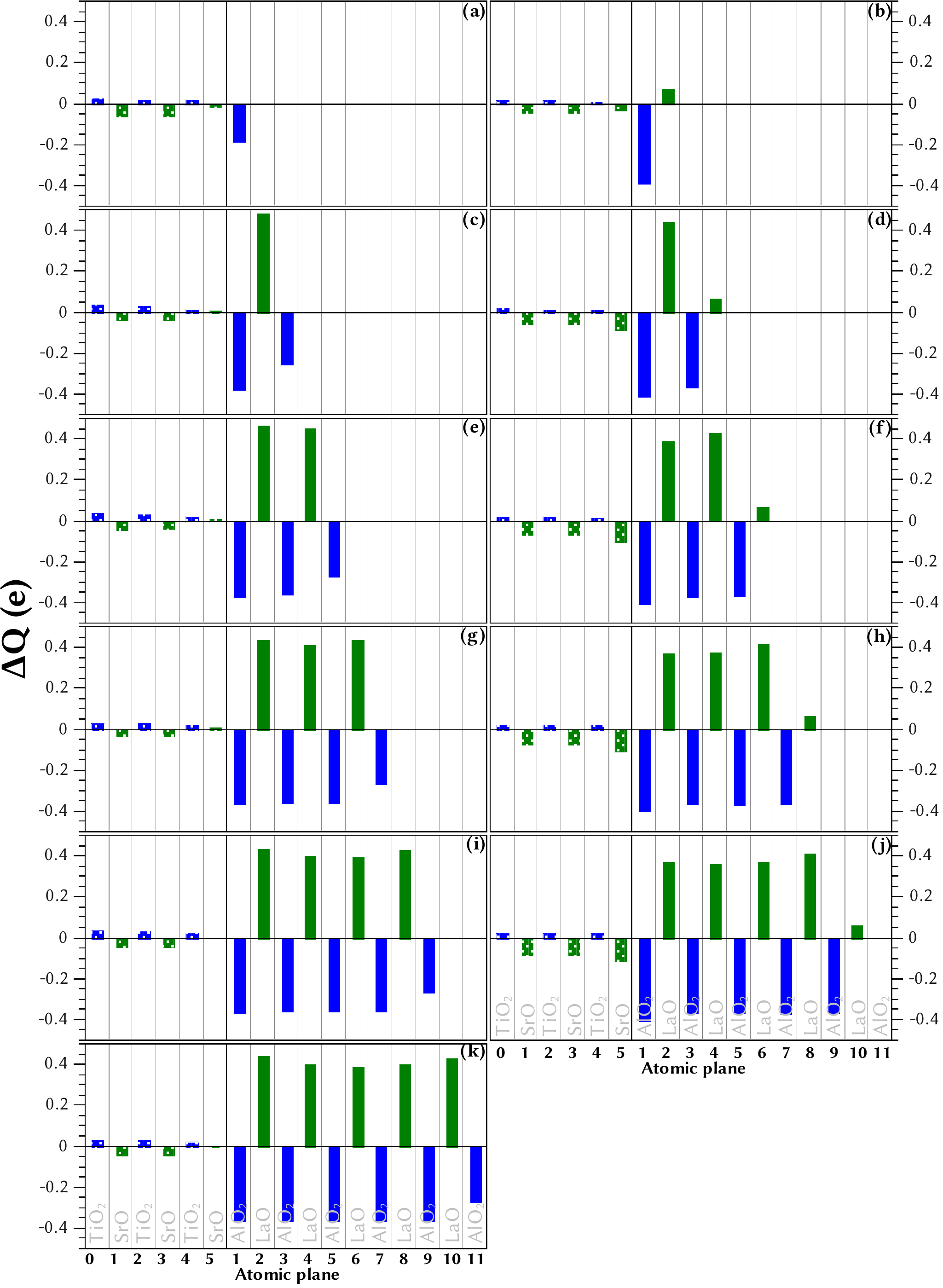}
\caption{\label{fig:Qpsupp} (Color online) Calculated deviations of Mulliken effective charge densities ($\Delta P_Q$) in AO- and BO$_2$-monolayers of p-LAO/STO(001) heterostructures with respect to charge densities in AO- and BO$_2$-monolayers of STO and LAO bulk, correspondingly. Calculations are performed using B3PW hybrid exchange-correlation functional. The $x$ axis shows the atomic monolayers from which atoms are originated. STO and LAO monolayers are numbered starting from the center of slab (0 means the central monolayer of the symmetrical slab unit cell). Monolayers (planes) are numbered separately for STO(001) substrate and for LAO(001) nanofilm. Number of monolayers increases from 1 to 11 for panels (a) to (k).}
\end{figure*}
\end{document}